\documentclass[showpacs,preprint,aps,superscriptaddress,floatfix,nofootinbib]{revtex4}
\usepackage{amsmath}
\usepackage[dvips]{graphicx,epsfig,color}
\newcommand{\oxygen}{{${}^{16}$O}}
\newcommand{\carbon}{{${}^{12}$C}}
\newcommand{\carbonn}{{${}^{13}$C}}
\newcommand{\calcium}{{${}^{48}$Ca}}
\definecolor{grey}{rgb}{0.93,0.93,0.93}

\begin{document}

\title{Spin-Excitation Mechanisms in Skyrme-Force Time-Dependent Hartree-Fock}
\author{J.~A.~Maruhn}
\affiliation{Institut f\"ur Theoretische Physik, Universit\"at Frankfurt, 
Max-von-Laue-Str. 1, 60438 Frankfurt am Main, Germany}
\author{P.-G.~Reinhard}
\affiliation{Institut f\"ur Theoretische Physik II, Universit\"at
  Erlangen-N\"urnberg, Staudtstrasse 7, D-91058 Erlangen, Germany}
\author{P.~D.~Stevenson}
\affiliation{Department of Physics, University of Surrey, Guildford,
  Surrey, GU2 7XH, United Kingdom}
\author{M.~R.~Strayer}
\affiliation{Physics Division, Oak Ridge National Laboratory, Oak Ridge, TN 37831-6373, USA}
\date{\today}
\begin{abstract}
We investigate the role of odd-odd (with respect to time inversion) couplings in the Skyrme force on
collisions of light nuclei, employing a fully three-dimensional numerical treatment without any
symmetry restrictions and with modern Skyrme functionals. We demonstrate the necessity of these
couplings to suppress spurious spin excitations owing to the spin-orbit force in free translational
motion of a nucleus but show that in a collision situation there is a strong spin excitation even in
spin-saturated systems which persists in the departing fragments. The energy loss is considerably
increased by the odd-odd terms.
\end{abstract}

\maketitle

Time-dependent Hartree--Fock (TDHF) enjoyed a period of large attention in nuclear physics about
thirty years ago; for reviews see e.g. \cite{Sve79a,Neg82aR,Dav85a}. These early calculations
delivered a great number of useful insights into the basic mechanisms of heavy-ion collisions, even
with the large practical restrictions of that time concerning the model, degrees of freedom, and
symmetries. However, it was soon recognized not to be as comprehensive a description as originally
expected. For example, widths in the distributions of fragments and kinetic energies are
systematically underestimated, a fact which had been traced back in parts to missing correlations
\cite{Rei83a,Mar85d}. More puzzling was that average quantities, such
as fusion cross sections, did not come out all that well although they should be predictable by
mean field dynamics. Already at that time there were indications that the many restrictions in the
calculations spoil their predictive value and that, for example, simply the proper handling of the
spin-orbit (l*s) force can improve the results considerably towards the experimentally observed
dissipation \cite{Uma86a,Rei88a}. Computer limitations halted these developments for a while. The
subsequent dramatic advance in computational power now allows three-dimensional TDHF calculations
with a full-fledged Skyrme force, without any symmetry restrictions, and for any nuclear size.
Accordingly, there is a renewed interest in TDHF studies as seen from recent publications on
resonance dynamics \cite{Nak05a,Rei05c,Sim03a} and heavy-ion collisions \cite{Uma06a}.  The present
manuscript also deals with recent 3D TDHF calculations and aims to investigate the importance of a
full treatment of the l*s force and related dissipation mechanisms.

TDHF in a nuclear context means a time-dependent mean-field theory derived from an effective energy
functional. Most widely used is the Skyrme functional which was proposed long ago as a quantitative
self-consistent model for the nuclear ground state \cite{Vau72a} and dynamics \cite{Eng75a}.  The
Skyrme energy-density functional consists of free kinetic energy, Coulomb energy with exchange in
the Slater approximation, and an effective-interaction part depending on density $\rho$, kinetic
density $\tau$, l*s density $\vec{J}$, current $\vec\jmath$, and spin density $\vec\sigma$, for a
detailed explicit expression see e.g. \cite{Ben03aR}.  Pairing is not considered in the present case
where we deal mostly with closed shell nuclei.  For the purpose of later discussions, we display
here the l*s part of the functional
\begin{eqnarray}
  E_{\rm ls}^{\rm(even)}
 \! &=&\!
     -\!\int\!d^3r\,\big( b_4\rho\nabla\!\cdot\!\vec{{J}}
     +b'_4\sum_q  \rho_q(\nabla\!\cdot\!\vec{J}_q)\big),
\label{eq:lseven}\\
  {E}_{\rm ls}^{\rm(odd)}
  \!&=&\!
  -\!\int\!{d}^3r
  \big(b_4\vec\sigma\!\cdot\!(\nabla\!\times\!{\vec\jmath})
  +b_4'\sum_q\vec\sigma_q\!\cdot\!(\nabla\!\times\!{\vec\jmath}_q)\big).
  {\rm~~}
\label{eq:lsodd}
\end{eqnarray}
The index $q\in \{p,n\}$ labels protons and neutrons. The specification of the energy-density
functional fixes all that is needed for TDHF and the stationary initial states.  The TDHF equations
are derived by time-dependent variation with respect to the single-nucleon wavefunctions
$\varphi^+_{\alpha}$ and the corresponding stationary HF equation by analogous stationary variation.

The full Skyrme functional and the subsequent TDHF equations meet all
symmetries of space-time, in particular invariance under Galilei
transformations, a condition which must be fulfilled for a meaningful theory
of heavy-ion reactions \cite{Eng75a}. Galileian invariance imposes restrictions on the form of the
odd-odd terms, i.e. those terms containing the time-odd pieces, current
$\vec\jmath$ and spin density $\vec\sigma$.  This means that the kinetic term
always appears in the boost-invariant combination $\rho\tau-\vec\jmath^2$.  Of
particuar importance here is the correct interplay between the even-even
and odd-odd parts of the l*s term, eqs. (\ref{eq:lseven}) with
(\ref{eq:lsodd}). Studies of rotational states confirmed
the large influence of the time-odd pieces, but concluded that better
agreement with experiment is obtained when some or all of these are omitted
\cite{Dob95c,Mol00a}.  We shall now discuss their effect on heavy-ion
collisions and, in particular, show that their omission leads to
inconsistencies. 

The Skyrme functional allows a very precise description of nuclear ground state properties and
excitations \cite{Ben03aR}.  There exists, in fact, a great variety of parametrizations of the
Skyrme functional in the literature which differ in quality and bias of fitted data.  In order to
distinguish generic effects from particularities of a certain parametrization, we considered several
different Skyrme forces and show results for two: SkM* \cite{Bar82a} and SLy6
\cite{Cha97a}. Calculations with other forces did not differ significantly for the purposes of this
work.

The practical solution of the TDHF equations employs a representation of wavefunctions, potentials,
and densities on a three-dimensional Cartesian coordinate-space grid. Derivatives are evaluated in
Fourier-transformed space using the fast Fourier transformation (FFT) \cite{Blu92a}. We work with a
grid spacing of 1 fm. The accelerated gradient iteration is employed to find the stationary ground
state solution \cite{Blu92a,Rei82a}.  The Coulomb field is calculated by solving the Poisson
equation on a grid which is twice as large as the physical grid and with periodic boundary
conditions with the method of \cite{Eas79a}. Note that the reflection of emitted nucleons from the
boundaries of the numerical box lead to an uncertainty of 2-3~MeV in the final relative-motion
energy, an effect which is also seen in giant resonance calculations \cite{Rei06a}.  A Taylor series
expansion up to sixth order of the unitary mean-field propagator is used for the dynamical time
stepping \cite{Flo78a}. The conservation of particle number and total energy provides a rather
stringent check of numerical accuracy. In practice, we tune our numerical parameters such that we
observe over all time a change in the particle number of less than 0.01, and a drift in the total
energy of less than 0.1~MeV. A time step of $\Delta t=0.2{\rm fm}/c$ was found adequate in all cases
considered, independent of the bombarding energy and also of whether the odd-odd l*s couplings were
included or not.

The standard test case throughout this paper is an \oxygen+\oxygen\ collision at $E_{\rm
cm}=$75--150~MeV. This system is one of the most frequently studied with TDHF and has also been the
focus of investigations of the effect of the l*s force on dissipation \cite{Rei88a}.  The fragment
wave functions are placed symmetrically on the grid to an initial c.~m. distance of 16~fm and then
boosted to the desired relative center-of-mass (c.m.)  energy. This prepares the initial state from
which the TDHF propagation is calculated. We also compare results for systems including \carbon,
\carbonn, and \calcium\ in order to get a first impression of the systematic variations.

The most interesting observable in heavy-ion collisions is the kinetic energy of relative motion of
the two fragments. This quantity is deduced using a two-body analysis of the time-dependent density
distribution. For this purpose, we calculate the principal axes of the mass quadrupole tensor, then
examine the density along the axis of minimum quadrupole moment to find whether it shows the
characteristics of two maxima separated by a low-density region. The point of lowest density along
this line then defines a dividing plane perpendicular to this axis, and two fragments are assumed to
exist on both sides of this plane. Calculating the centers of mass of each fragment yields a new
straight line connecting them, which is used to repeat the process. This is iterated until the
definitions of the fragment centers of mass and the dividing plane have stabilized. The principal
result of this analysis are the fragment masses and charges $M_i$, $Q_i$, $i=1,2$, the separation
distance $R$ of the fragments, the relative velocity $\dot R$ as well as the angular velocity of
rotation in the scattering plane $\dot\theta$ calculated from the positions in two successive time
steps.  The energy of relative motion can then be calculated from the radial kinetic and rotational
energy of the fragment minus the remaining Coulomb interaction.  The Coulomb energy is approximated
by the expression for two point charges, which should be good for larger distances. By comparing
this with the full numeric calculation of the Coulomb interaction energy, we could establish that
for $R\geq12$~fm it is accurate within about 0.02~MeV.
\begin{figure}
\begin{center}
\includegraphics[width=7.0cm]{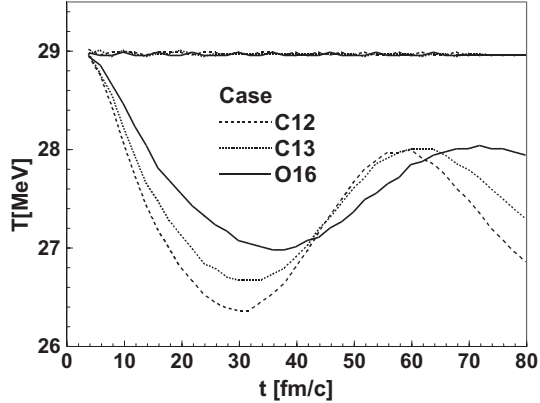}
\end{center}
\caption{The c.m. energy of a single free nucleus moving through the grid
for three different test cases as inidicated.  The nearly
constant lines correspond to the full Skyrme treatment, while the
other results were calculated omitting the odd-odd l*s coupling.
\label{fig:transv}}
\end{figure}

As a first critical test, we consider the free translational motion of a
nucleus.  The results for three different nuclei shown in Fig.~\ref{fig:transv}
strikingly illustrate, on the one hand, the accuracy of the code, and on the
other, the need for odd-odd couplings. If the full Skyrme force is used, the
kinetic energy of the c.m.\ just shows small oscillations by $|\Delta T|<0.2$~MeV 
in the period covered, while with the
odd-odd couplings omitted there is an immediate deceleration followed
by large oscillations. At this point we can already conclude
that {\em the omission of these terms leads to unacceptable physical
behaviour}.

The spurious dissipation is caused by the intrinsic excitation of a
spin-twist mode.  For a nucleus moving with constant velocity $\vec v$, the
coupling term contains
\begin{equation}
  \nabla\!\times\!\vec\jmath=(\nabla\rho)\!\times\!\vec v=\frac{{\rm
      d}\rho}{{\rm d}r}\frac{\vec r\!\times\!\vec v}{r},
\label{eq:coupl}
\end{equation}
where spherical symmetry was assumed for simplicity. This is an
azimuthal vector field which thus couples to a spin field of the same
character. {\em Omitting the odd-odd coupling thus leads to a spurious
  excitation of a ring-like spin density}, which with the full Skyrme
force is suppressed by the odd-odd terms.
An examination of the spin density in the cases without the odd-odd l*s terms
shows that the actual excitation of this mode accounts for about 95\% of the
energy loss. The rest is due to additional excitations caused by the
deceleration (note that the total energy is conserved in any case).
\begin{figure}
\includegraphics[width=8.0cm]{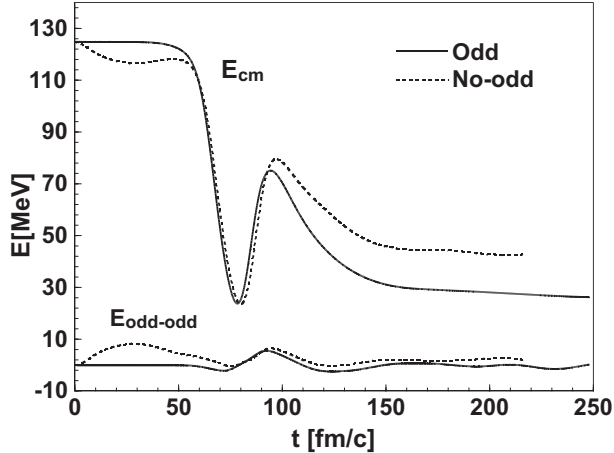}
\caption{\label{fig:einit} 
Relative c.m. energy (top) and odd-odd l*s
energy (bottom) in a head-on \oxygen+\oxygen\ collision at 125~MeV. The
energies values loose their meaning in the contact regime, which is between
about 50 and 120~fm/c. ``odd'' and ``no-odd'' refer to calculation with and
without the odd-odd l*s couplings.}
\end{figure}

Fig.~\ref{fig:einit} shows the relative c.m. energy for a collision.  The initial phase (up to 40
fm/c) is free c.m. motion of the two nuclei, and we see again the spurious dissipation of
c.m. energy as soon as the odd-odd l*s term is omitted.  In contrast, the full Skyrme interaction
preserves the relative motion energy very well until contact.  The bottom part of the figure shows
the energy contained in the odd-odd l*s term (\ref{eq:lsodd}). The ``no-odd'' case shows a
substantial increase in that energy which is obviously properly compensated in the full
treatment. The energetic relations are reversed in the exit channel. The unphysical case without
odd-odd l*s departs with more residual c.m. energy while the full interaction produces more true
dissipation.  This is caused again by the mechanism sketched in the spin-coupling term
(\ref{eq:coupl}), but {\em now is not spurious:} it occurs because the l*s
terms in the single-particle Hamiltonian and the counterbalancing odd-odd l*s coupling come out of
synchronization due to the physical change in current pattern during the collision, so that a net
spin-twist excitation remains.
\begin{figure}
\begin{center}
\includegraphics[width=4.5cm]{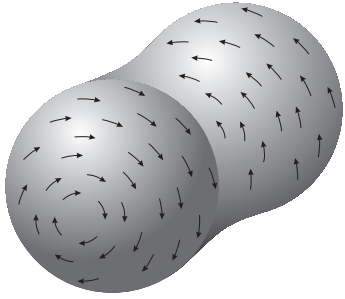}
\end{center}
\caption{\label{spinex}
Artist's view of the spin excitation generated in a central
collision of two \oxygen\ nuclei. Closely based on the numerical spin-density vectors produced in
the calculations.}
\end{figure}
Fig.~\ref{spinex} visualizes the pattern of the spin-twist excitation in a stage where the two
fragments start interacting. This spin density distribution pattern was observed in the numerical
results. The excitation persists after full separation. The extra amount of energy stored in this
mode explains the enhanced dissipation observed in the exit stages of Fig.~\ref{fig:einit}.  The
spin-twist mode and the unambigous detection of its dexcitation following a collision certainly
deserves further detailed investigations.

\begin{figure}
\includegraphics[width=8.8cm]{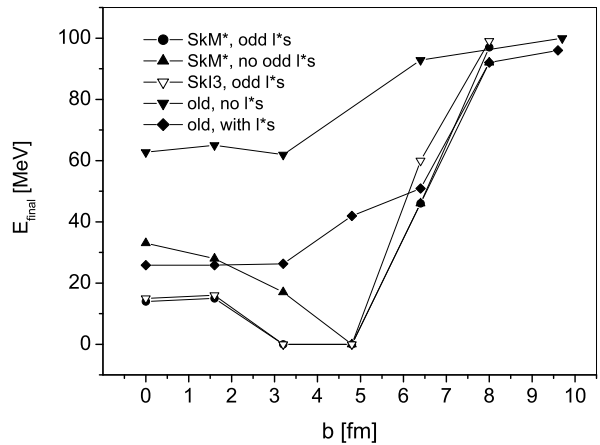}
\caption{\label{fig:noncentral}
The final c.~m.\ kinetic energies in noncentral collisions of
\oxygen\ on \oxygen\ at 100~MeV. at the impact parameters given, and
with or without the inclusion of the odd-odd $l*s$ terms.  The Skyrme
force used was SkM*. For comparison, results from Ref.~\cite{Rei88a}
are shown, with and (completely) without $l*s$-force.}
\end{figure}

We have investigated a great variety of central and non-central collisions at various collisional
energies. They all show similar effects.  As one example, Fig.~\ref{fig:noncentral} shows the loss
of c.~m. energy versus impact parameter, mainly for SkM* at various levels of approximation, one
result from SkI3 for comparison, and results from older TDHF studies in axial approximation
\cite{Uma86a,Rei88a}.  The old calculations were done with a variant
of SkM* replacing the gradient terms by finite folding. At that time,
it was great success to include the even-even l*s term. This brought a
substantial jump in dissipation and resolved the puzzle of too much
transparency in the then older TDHF calculations. The present
calculations without odd-odd terms still differ from the previous ones
in that they are now fully triaxial. This makes no effect at low impact
parameter (the minor difference is probably due to the folding
approximation) but a visibly enhanced dissipation for grazing
collisions which is reasonable because non-central collisions
break axial symmetry and call for a triaxial treatment. The most
interesting effect here is the additional dissipation caused by the
step to the full Skyrme functional (compare up-triangles with full
dots). It remains very similar up to an impact
parameter of to about 5~fm which is, not suprisingly, close to the
nuclear radius. For larger $b$, it rapidly vanishes as we get to
peripheral collisions. The spin-twist mode thus leads to excitations
in the final fragments that remain relevant for a large range of
impact parameters.
The most striking effect is the qualitative difference that thew new calculations without symmetry
restrictions and with all odd-odd terms included predict a large regime of fusion whereas all the
other do not. We can estimate a fusion cross section for the system $^{16}$O+$^{16}$O at collisional
energy of 100 MeV to be somewhat larger than 500 mb. This compares very well with 550 mb deduced
from the systematics of \cite{Bir79a} while all approximate calculations fail in that respect.
Fig.~\ref{fig:noncentral} also shows the result for full SkI3, which is very similar to SkM*. The
same similarity is seen for other forces investigated.  Calculations with older Skyrme forces showed
a much more dramatic and systematic force dependence \cite{Mar85a,Uma86a,Rei88a,Uma89}.

The dependence of the additional dissipation (as compared to restricted
calculations) on the mass of the colliding nuclei was tested by a few
calculations for other collision partners. For \carbon\ on \carbon\ the
additional energy loss was similar and possibly even larger reaching 30~MeV, and
similarly for \carbon\ on \oxygen. On the other hand, for \calcium\ on
\calcium\ it was almost negligible with about 1~MeV. The reason for this
is not yet clear and needs to be investigated.  

In this work we have investigated the energy loss in
heavy-ion collisions as described by TDHF handled in full 3D and including consistently all terms of
the given Skyrme functional. Particular attention was paid to the time reversal odd-odd spin-orbit
(l*s) term which is often neglected in TDHF calcuations. The main findings can be concluded in
brief: The odd-odd l*s terms establish full Galileian invariance of the functional and they are
crucial to provide properly free translation of a nucleus over the grid.  The odd-odd l*s terms add
substantially to the dissipation observed in heavy-ion collisions. That effect persists up to impact
parameters of order of the nuclear radius. It is large for small nuclei and seems to decrease for
heavier ones.  The enhanced dissipation is associated with the strong excitation of a pronounced
spin-twist mode which is present even in the collision of spin-saturated nuclei and persists after
separation in both fragments.  The two main tasks for future research are: first, large scale
investigations of dissipation under varying scattering conditions, and second, a closer inspection
of that most interesting spin-twist mode, working out directions for an experimental assessment.

\noindent
This work was supported by BMBF under contracts no. 06 F 131 and 06 ER
808 and the UK EPSRC grant GR/96425/01. We gratefully acknowledge support by the
Frankfurt Center for Scientific Computing.

\bibliographystyle{apsrev}

\bibliography{tdhf}
\end{document}